\documentclass[preprint,showpacs,preprintnumbers,amsmath,amssymb,floatfix]{revtex4}
\usepackage{amsmath}

\usepackage{subfigure}
\usepackage{float}
\usepackage{graphicx}
\usepackage{dcolumn}
\usepackage{bm}
\usepackage{color}
\definecolor{orange}{RGB}{255,127,0}
\definecolor{brown}{RGB}{160,82,45}

\def\undersim#1{\setbox9\hbox{${#1}$}{#1}\kern-\wd9\lower
    2.5pt \hbox{\lower\dp9\hbox to \wd9{\hss $_\sim$\hss}}}
\def\undersim#1{\setbox9\hbox{${#1}$}{#1}\kern-\wd9\lower
    2.5pt \hbox{\lower\dp9\hbox to \wd9{\hss $_\sim$\hss}}}

\def\mr{{\mathbf r}}

\def\mr{{\mathbf r}}

\def\mk{{\mathbf p}}
\def\bk{{\mathbf k}}

\begin{document}
\vspace*{-10mm}
\title{Multiboson Hanbury Brown-Twiss correlations for partially coherent sources
in relativistic heavy-ion collisions in a multiphase transport model}

\author{Shi-Yao Wang$^1$, Jun-Ting Ye$^1$, Wei-Ning
Zhang$^{1,2}$\footnote{wnzhang@dlut.edu.cn}}
\affiliation{$^1$School of Physics, Dalian University of Technology, Dalian,
Liaoning 116024, China\\
$^2$School of Physics, Harbin Institute of Technology, Harbin, Heilongjiang
150006, China}


\begin{abstract}
We use a multi-phase transport (AMPT) model to study multi-pion and multi-kaon Hanbury
Brown-Twiss (HBT) correlations for the partially coherent particle-emitting sources in
relativistic heavy-ion collisions. A density-dependent longitudinal coherent emission
length and density-dependent transverse coherent emission length are introduced in
calculating the multi-boson HBT correlation functions of the partially coherent sources.
We compare the model results of three- and four-pion HBT correlation functions with
experimental data in Pb-Pb collisions at center-of-mass energy $\sqrt{s_{NN}}=$2.76 TeV,
and investigate the influences of boson coherent emissions on the multi-pion and multi-kaon correlation functions. We find that all of the three- and four-pion
correlation functions of the partially coherent sources are consistent with experimental
data. Coherent emission leads to the intercept decreases of the multi-boson correlation
functions. The intercepts of the multi-kaon correlation functions of the partially
coherent source are higher than those of the multi-pion correlation functions, because
low kaon densities lead to smaller kaon coherent emission lengths than pion emission
lengths. The intercepts of multi-boson correlation functions of partially coherent
sources in high transverse momentum intervals are higher than those in low transverse
momentum intervals because particle de Broglie wavelengths are small at high momenta.

\end{abstract}


\maketitle

\section{Introduction}
Identical boson Hanbury Brown-Twiss (HBT) correlations are important observables in
high-energy heavy-ion collisions \cite{Gyu79,Wongbook94,Wie99,Wei00,Csorgo02,Lisa05}.
At the Large Hadron Collider (LHC), the multiplicities of identical bosons in an
event are high in nucleus-nucleus collisions, e.g., several thousands for pions and
several hundreds for kaons \cite{ALICE_PRC88_2013}.
So, the study of multi-boson HBT correlations becomes possible
\cite{{ALICE_PLB739_2014,ALICE_PRC89_2014,ALICE_PRC93_2016,Gangadharan_PRC92_2015,
BaryRuZhang_JPG45_2018,BaryRuZhang_JPG46_2019,BaryZhangRuYang_CPC45_2021}}.
Pions are the most abundant boson produced in high-energy nucleus-nucleus collisions.
The observed suppressions of two- and multi-pion HBT correlation functions at small
relative momenta in experiments may indicate that the pion-emitting sources are
partially coherent \cite{{Gyu79,Wongbook94,Wie99,Wei00,Csorgo02,Lisa05,
ALICE_PLB739_2014,ALICE_PRC89_2014,ALICE_PRC93_2016,Gangadharan_PRC92_2015,
BaryRuZhang_JPG45_2018,BaryRuZhang_JPG46_2019,BaryZhangRuYang_CPC45_2021,
STAR_PRL87_2001,PHENIX_PRL93_2004,STAR_PRC71_2005,STAR_PRC81_2010,ALICE_PRC92_2015}}.

Compared with two-pion HBT correlations, multi-pion HBT correlations are sensitive
to the source coherence  \cite{{Liu_PRC_86,HeinzZhang_PRC_97,HeinzSugarbaker_PRC_04,
Gangadharan_PRC92_2015,BaryRuZhang_JPG45_2018,BaryRuZhang_JPG46_2019,
BaryZhangRuYang_CPC45_2021}}. In this paper, we study the pion and kaon three- and
four-particle HBT correlations for partially coherent sources in Pb-Pb central
collisions at $\sqrt{s_{NN}}=$2.76~TeV, using a multi-phase transport model (AMPT)
\cite{LinKo_PRC65_2002,Lin_PRC72_2005}. We construct the partially coherent sources
for pion and kaon emissions with a consistent longitudinal coherent emission length
and a transverse coherent emission length, which are dependent on the particle de
Broglie wavelengths \cite{WangYeZhang_PRC109_2024}, as well as the boson densities
at generation configurations in the AMPT model. It is found that both of the three-
and four-pion correlation functions of the partially coherent sources calculated in
the AMPT model are consistent with the experimental results in Pb-Pb collisions at
$\sqrt{s_{NN}}=$2.76~TeV at the LHC \cite{ALICE_PRC93_2016}. The intercepts of the
multi-boson correlation functions of partially coherent sources reflect the influences
of boson coherent emission on the correlation functions. We find that the intercepts
of the multi-kaon correlation functions of partially coherent sources are higher
than those of multi-pion correlation functions, and the intercepts of the multi-boson
correlation functions in high transverse momentum intervals are higher than those in
low transverse momentum intervals.

The rest of this paper is organized as follows. In Section II, we introduce the consistent
boson density dependent coherent emission lengths for the pion and kaon partially coherent
sources, and describe the calculation of the multi-boson HBT correlation functions in the
AMPT model. Section III presents the results of pion and kaon multi-particle HBT correlation
functions in Pb-Pb central collisions at $\sqrt{s_{NN}}=$2.76~TeV in the AMPT model, and
compares the model results of three- and four-pion correlation functions with experimental
data. We also calculate the intercepts of the multi-boson correlation functions of the
partially coherent sources, and investigate the influences on them of boson coherent
emission. We summarize this work and present conclusions in Section IV.

\section{Multiboson HBT correlations for partially coherent sources in the AMPT model}

\subsection{Multiboson HBT correlation functions}
The HBT correlation functions of $m\,(\ge2)$ bosons are defined as
\begin{equation}
C_m(\mk_1,\mk_2,\cdots,\mk_m)=\frac{P_m(\mk_1,\mk_2,\cdots,\mk_m)}{P_1(\mk_1)P_1(\mk_2)
\cdots P_1(\mk_m)},
\label{Cm1}
\end{equation}
where $P_1(\mk_i)~(i=1,2,\cdots,m)$ is the distribution of single-boson momentum $\mk_i$,
and $P_m(\mk_1,\mk_2,\cdots,\mk_m)$ is the $m$ identical boson momentum distribution in
an event.

Considering the bosons emitted from completely chaotic sources and with small momentum
differences, we can write the denominator and numerator, respectively, in Eq.~(\ref{Cm1})
in the smoothed approximation as \cite{WangYeZhang_PRC109_2024}
\begin{equation}
P_1(\mk_1)P_1(\mk_2)\cdots P_1(\mk_m)=\sum_{X_1}\sum_{X_2}\cdots\sum_{X_m} A^2(\mk_1,X_1)
A^2(\mk_2,X_2)\cdots A^2(\mk_m,X_m),
\label{P1m}
\end{equation}
\begin{eqnarray}
&&P_m(\mk_1,\mk_2,\cdots\mk_m)=\sum_{X_1}\sum_{X_2}\cdots\sum_{X_m} A^2(\mk_1,X_1)
A^2(\mk_2,X_2)\cdots A^2(\mk_m,X_m) \cr
&&\hspace*{36mm}\times \Big|\Psi(\mk_1, \mk_2,\cdots,\mk_m;
X_1,X_2,\cdots,X_m)\Big|^2,
\label{Pm1}
\end{eqnarray}
where $A(\mk_i,X_i)$ is the magnitude of the amplitude for emitting a pion with
momentum $\mk_i$ at four-coordinate $X_i$ (freeze-out coordinates), and
\begin{equation}
\Psi(\mk_1, \mk_2,\cdots,\mk_m;X_1,X_2,\cdots,X_m)=\frac{1}{\sqrt{m!}}\sum_{\sigma}
\sum_{j=1}^{m} e^{ip_j\cdot X_{\sigma(j)}},
\end{equation}
where $p_j$ is the four-momentum for $\mk_j$, $\sigma(j)$ is the $j$th element
of a permutation of the sequence $\{1,2, \cdots,m\}$, and $\sum_{\sigma}$ denotes
summation over all $m!$ permutations of the sequence.

For $m=$2, 3, and 4, respectively, we have
\begin{eqnarray}
&&C_2(\mk_1,\mk_2)=\sum_{X_1}\sum_{X_2} A^2(\mk_1,X_1)A^2(\mk_2,X_2) \Big\{1+{\rm Re}
\big[f(p_1-p_2,X_1)f^*(p_1-p_2,X_2)\big]\Big\}\cr
&&\hspace*{22mm}\bigg/\sum_{X_1}\sum_{X_2} A^2(\mk_1,X_1)A^2(\mk_2,X_2)\cr
&&\hspace*{16mm}=\sum_{X_1}\sum_{X_2} A^2(\mk_1,X_1)A^2(\mk_2,X_2) \Big\{1+\cos
\big[(p_1-p_2)\cdot(X_1-X_2)\big]\Big\}\cr
&&\hspace*{22mm}\bigg/\sum_{X_1}\sum_{X_2} A^2(\mk_1,X_1)A^2(\mk_2,X_2),
\label{C21}
\end{eqnarray}
\begin{eqnarray}
&&C_3(\mk_1,\mk_2,\mk_3)=\sum_{X_1}\sum_{X_2}\sum_{X_3} A^2(\mk_1,X_1)A^2(\mk_2,X_2)
A^2(\mk_3,X_3)\cr
&&\hspace*{10mm}\times \Big\{1+\!{\rm Re}\big[f(p_1-p_2,X_1)f^*(p_1-p_2,X_2)\big]
+\!{\rm Re}\big[f(p_1-p_3,X_1)f^*(p_1-p_3,X_3)\big]\cr
&&\hspace*{19mm}+{\rm Re}\big[f(p_2-p_3,X_2)f^*(p_2-p_3,X_3)\big]\cr
&&\hspace*{19mm}+2{\rm Re}\big[f(p_1-p_2,X_1)f(p_2-p_3,X_2)f(p_3-p_1,X_1)\big]\Big\}\cr
&&\hspace*{10mm}\bigg/\sum_{X_1}\sum_{X_2}\sum_{X_3} A^2(\mk_1,X_1)A^2(\mk_2,X_2)
A^2(\mk_3,X_3),
\label{C31}
\end{eqnarray}
\begin{eqnarray}
&&C_4(\mk_1,\mk_2,\mk_3,\mk_4)=\sum_{X_1}\sum_{X_2}\sum_{X_3}\sum_{X_4} A^2(\mk_1,X_1)
A^2(\mk_2,X_2)A^2(\mk_3,X_3)A^2(\mk_4,X_4)\cr
&&\hspace*{10mm}\times \Big\{1+\!{\rm Re}\big[f(p_1-p_2,X_1)f^*(p_1-p_2,X_2)\big]
+\!{\rm Re}\big[f(p_1-p_3,X_1)f^*(p_1-p_3,X_3)\big]\cr
&&\hspace*{19mm}+{\rm Re}\big[f(p_1-p_4,X_1)f^*(p_1-p_4,X_4)\big]
+{\rm Re}\big[f(p_2-p_3,X_2)f^*(p_2-p_3,X_3)\big]\cr
&&\hspace*{19mm}+{\rm Re}\big[f(p_2-p_4,X_2)f^*(p_2-p_4,X_4)\big]
+{\rm Re}\big[f(p_3-p_4,X_3)f^*(p_3-p_4,X_4)\big]\cr
&&\hspace*{19mm}+2{\rm Re}\big[f(p_1-p_2,X_1)f(p_2-p_3,X_2)f(p_3-p_1,X_3)\big]\cr
&&\hspace*{19mm}+2{\rm Re}\big[f(p_1-p_2,X_1)f(p_2-p_4,X_2)f(p_4-p_1,X_4)\big]\cr
&&\hspace*{19mm}+2{\rm Re}\big[f(p_1-p_3,X_1)f(p_3-p_4,X_3)f(p_4-p_1,X_1)\big]\cr
&&\hspace*{19mm}+2{\rm Re}\big[f(p_2-p_3,X_2)f(p_3-p_4,X_3)f(p_4-p_2,X_4)\big]\cr
&&\hspace*{19mm}+{\rm Re}\big[f(p_1-p_2,X_1)f^*(p_1-p_2,X_2)\big]
{\rm Re}\big[f(p_3-p_4,X_3)f^*(p_3-p_4,X_4)\big]\cr
&&\hspace*{19mm}+{\rm Re}\big[f(p_1-p_3,X_1)f^*(p_1-p_3,X_3)\big]
{\rm Re}\big[f(p_2-p_4,X_2)f^*(p_2-p_4,X_4)\big]\cr
&&\hspace*{19mm}+{\rm Re}\big[f(p_1-p_4,X_1)f^*(p_1-p_4,X_4)\big]
{\rm Re}\big[f(p_2-p_4,X_2)f^*(p_2-p_3,X_3)\big]\cr
&&\hspace*{19mm}+2{\rm Re}\big[f(p_1-p_2,X_1)f(p_2-p_3,X_2)f(p_3-p_4,X_3)
f(p_4-p_1,X_4)\big]\cr
&&\hspace*{19mm}+2{\rm Re}\big[f(p_1-p_2,X_1)f(p_2-p_4,X_2)f(p_4-p_3,X_4)
f(p_3-p_1,X_3)\big]\cr
&&\hspace*{19mm}+2{\rm Re}\big[f(p_1-p_3,X_1)f(p_3-p_2,X_3)f(p_2-p_4,X_2)
f(p_4-p_1,X_4)\big]\Big\}\cr
&&\hspace*{10mm}\bigg/\sum_{X_1}\sum_{X_2}\sum_{X_3}\sum_{X_4} A^2(\mk_1,X_1)
A^2(\mk_2,X_2)A^2(\mk_3,X_3)A^2(\mk_4,X_4),
\label{C41}
\end{eqnarray}
where
\begin{equation}
f(p_i-p_j,X)=e^{i(p_i-p_j)\cdot X}=f^*(p_j-p_i,X),~~~~i,j=1,2,3,4,i\neq j,
\end{equation}
is called the amplitude of the correlator between the $i$th and $j$th bosons.

In Eq.~(\ref{C31}) the last term in curly braces expresses the pure triplet
correlation of the three bosons. The three-boson cumulant correlation function
is defined as
\begin{eqnarray}
&&c_3(\mk_1,\mk_2,\mk_3)=\sum_{X_1}\sum_{X_2}\sum_{X_3} A^2(\mk_1,X_1)A^2(\mk_2,X_2)
A^2(\mk_3,X_3)\cr
&&\hspace*{10mm}\times \Big\{1+2{\rm Re}\big[f(p_1-p_2,X_1)f(p_2-p_3,X_2)f(p_3-p_1,X_1)
\big]\Big\}\cr
&&\hspace*{10mm}\Big/ \sum_{X_1}\sum_{X_2}\sum_{X_3} A^2(\mk_1,X_1)A^2(\mk_2,X_2)
A^2(\mk_3,X_3).
\label{c31}
\end{eqnarray}

In Eq.~(\ref{C41}), the eighth through eleventh terms in curly braces express the
pure triplet correlations of the three bosons, the twelfth through fourteenth terms
in curly braces express the correlations of the double boson pairs, and the last
three terms in curly brackets express the pure quadruplet correlations of the four
bosons.
The four-boson cumulant correlation functions $c_4$, $a_4$, and $b_4$ are defined
respectively as \cite{ALICE_PRC93_2016,BaryRuZhang_JPG45_2018}
\begin{eqnarray}
&&c_4(\mk_1,\mk_2,\mk_3,\mk_4)=\sum_{X_1}\sum_{X_2}\sum_{X_3}\sum_{X_4} A^2(\mk_1,X_1)
A^2(\mk_2,X_2)A^2(\mk_3,X_3)A^2(\mk_4,X_4)\cr
&&\hspace*{10mm}\times \Big\{1
+2{\rm Re}\big[f(p_1-p_2,X_1)f(p_2-p_3,X_2)f(p_3-p_4,X_3)
f(p_4-p_1,X_4)\big]\cr
&&\hspace*{19mm}+2{\rm Re}\big[f(p_1-p_2,X_1)f(p_2-p_4,X_2)f(p_4-p_3,X_4)
f(p_3-p_1,X_3)\big]\cr
&&\hspace*{19mm}+2{\rm Re}\big[f(p_1-p_3,X_1)f(p_3-p_2,X_3)f(p_2-p_4,X_2)
f(p_4-p_1,X_4)\big]\Big\}\cr
&&\hspace*{10mm}\Big/ \sum_{X_1}\sum_{X_2}\sum_{X_3}\sum_{X_4}
A^2(\mk_1,X_1)A^2(\mk_2,X_2)A^2(\mk_3,X_3)A^2(\mk_4,X_4),
\label{c41}
\end{eqnarray}
\begin{eqnarray}
&&a_4(\mk_1,\mk_2,\mk_3,\mk_4)=\sum_{X_1}\sum_{X_2}\sum_{X_3}\sum_{X_4} A^2(\mk_1,X_1)
A^2(\mk_2,X_2)A^2(\mk_3,X_3)A^2(\mk_4,X_4)\cr
&&\hspace*{10mm}\times \Big\{1
+2{\rm Re}\big[f(p_1-p_2,X_1)f(p_2-p_3,X_2)f(p_3-p_1,X_3)\big]\cr
&&\hspace*{19mm}+2{\rm Re}\big[f(p_1-p_2,X_1)f(p_2-p_4,X_2)f(p_4-p_1,X_4)\big]\cr
&&\hspace*{19mm}+2{\rm Re}\big[f(p_1-p_3,X_1)f(p_3-p_4,X_3)f(p_4-p_1,X_1)\big]\cr
&&\hspace*{19mm}+2{\rm Re}\big[f(p_2-p_3,X_2)f(p_3-p_4,X_3)f(p_4-p_2,X_4)\big]\cr
&&\hspace*{19mm}+{\rm Re}\big[f(p_1-p_2,X_1)f^*(p_1-p_2,X_2)\big]
{\rm Re}\big[f(p_3-p_4,X_3)f^*(p_3-p_4,X_4)\big]\cr
&&\hspace*{19mm}+{\rm Re}\big[f(p_1-p_3,X_1)f^*(p_1-p_3,X_3)\big]
{\rm Re}\big[f(p_2-p_4,X_2)f^*(p_2-p_4,X_4)\big]\cr
&&\hspace*{19mm}+{\rm Re}\big[f(p_1-p_4,X_1)f^*(p_1-p_4,X_4)\big]
{\rm Re}\big[f(p_2-p_4,X_2)f^*(p_2-p_3,X_3)\big]\cr
&&\hspace*{19mm}+2{\rm Re}\big[f(p_1-p_2,X_1)f(p_2-p_3,X_2)f(p_3-p_4,X_3)
f(p_4-p_1,X_4)\big]\cr
&&\hspace*{19mm}+2{\rm Re}\big[f(p_1-p_2,X_1)f(p_2-p_4,X_2)f(p_4-p_3,X_4)
f(p_3-p_1,X_3)\big]\cr
&&\hspace*{19mm}+2{\rm Re}\big[f(p_1-p_3,X_1)f(p_3-p_2,X_3)f(p_2-p_4,X_2)
f(p_4-p_1,X_4)\big]\Big\}\cr
&&\hspace*{10mm}\Big/ \sum_{X_1}\sum_{X_2}\sum_{X_3}\sum_{X_4}
A^2(\mk_1,X_1)A^2(\mk_2,X_2)A^2(\mk_3,X_3)A^2(\mk_4,X_4),
\label{a41}
\end{eqnarray}
and
\begin{eqnarray}
&&b_4(\mk_1,\mk_2,\mk_3,\mk_4)=\sum_{X_1}\sum_{X_2}\sum_{X_3}\sum_{X_4} A^2(\mk_1,X_1)
A^2(\mk_2,X_2)A^2(\mk_3,X_3)A^2(\mk_4,X_4)\cr
&&\hspace*{10mm}\times \Big\{1
+2{\rm Re}\big[f(p_1-p_2,X_1)f(p_2-p_3,X_2)f(p_3-p_1,X_3)\big]\cr
&&\hspace*{19mm}+2{\rm Re}\big[f(p_1-p_2,X_1)f(p_2-p_4,X_2)f(p_4-p_1,X_4)\big]\cr
&&\hspace*{19mm}+2{\rm Re}\big[f(p_1-p_3,X_1)f(p_3-p_4,X_3)f(p_4-p_1,X_1)\big]\cr
&&\hspace*{19mm}+2{\rm Re}\big[f(p_2-p_3,X_2)f(p_3-p_4,X_3)f(p_4-p_2,X_4)\big]\cr
&&\hspace*{19mm}+2{\rm Re}\big[f(p_1-p_2,X_1)f(p_2-p_3,X_2)f(p_3-p_4,X_3)
f(p_4-p_1,X_4)\big]\cr
&&\hspace*{19mm}+2{\rm Re}\big[f(p_1-p_2,X_1)f(p_2-p_4,X_2)f(p_4-p_3,X_4)
f(p_3-p_1,X_3)\big]\cr
&&\hspace*{19mm}+2{\rm Re}\big[f(p_1-p_3,X_1)f(p_3-p_2,X_3)f(p_2-p_4,X_2)
f(p_4-p_1,X_4)\big]\Big\}\cr
&&\hspace*{10mm}\Big/ \sum_{X_1}\sum_{X_2}\sum_{X_3}\sum_{X_4}
A^2(\mk_1,X_1)A^2(\mk_2,X_2)A^2(\mk_3,X_3)A^2(\mk_4,X_4).
\label{b41}
\end{eqnarray}

\subsection{Calculation of multi-boson correlation functions in the AMPT model}
The AMPT model has been successfully used to describe the observables in high-energy
heavy-ion collisions
\cite{{LinKo_PRC65_2002,LinKo_JPG30_2004,Lin_PRC72_2005,Nasim_PRC82_2010,
XJunCMKo_PRC83_2011,DSolanki_PLB720_2013,YLXie_NPA920_2013,BzdakGLMa_PRL113_2014,
GLMaZWLin_PRC93_2016,HLi_PRC96_2017,Haque_JPG46_2019,MDordevic_PRC101_2020,
TShao_PRC102_2020,KJSunCMKo_PRC103_2021,MagdtLacey_PRC104_2021,SBasu_PRC104_2021,
MagEvdLac_JPG48_2021,ZZhangNYuHXu_EPJA58_2022,WangYeZhang_PRC109_2024}}.
It is a hybrid composed of initialization, parton transport, hadronization, and
hadron transport \cite{Lin_PRC72_2005}. The initialization of collisions in the
AMPT model is performed using the HIJING model \cite{HIJING}. The parton and hadron
transport are described by the ZPC (Zhang's parton cascade) model \cite{ZPC} and
ART model \cite{ART}, respectively.
In this study, we investigate the multi-boson correlations of identical pions
and kaons generated with the string melting version of the AMPT model
\cite{LinKo_PRC65_2002,Lin_PRC72_2005}, for Pb-Pb central collisions at
$\sqrt{s_{NN}}=2.76$~TeV. The impact parameter $b$ is taken between 0 and 3.5
fm for centrality 0--5\% \cite{ALICE_PRC88_2013a}, and the freeze-out times of the
pion and kaon sources are taken to be 40~fm/$c$. We take the model parameter $\mu$
of parton screening mass to be 2.2814~fm$^{-1}$ \cite{Lin_PRC72_2005} in the
calculations. The strong coupling constant $\alpha_s$ is taken to be 0.47,
corresponding to a parton-scattering cross section of 6~mb \cite{Lin_PRC72_2005}.

With the AMPT model, one can trace back to the origin of a freeze-out particle, its
generation coordinate $\mr=(x,y,z)=(\mr_{\rm T},z)$, momentum $\bk=(\bk_{\rm T},k_Z)$,
and parent.
We previously calculated the two-pion HBT correlation functions for the partially
coherent sources in the AMPT model~\cite{WangYeZhang_PRC109_2024}. It was assumed that
the emissions of two pions are coherent if their longitudinal difference of generation
coordinates $\Delta z=|z_1-z_2|$ is smaller than the longitudinal coherent emission
length $L_{\rm CZ}(k_{1Z},k_{2Z})$ and their transverse difference of generation
coordinates $\Delta r_{\rm T}=\sqrt{(x_1-x_2)^2+(y_1-y_2)^2}$ is smaller than the
transverse coherent emission length $L_{\rm CT}(k_{1\rm T},k_{2\rm T})$.
In this case, the contribution of this pair to the two-pion correlator in Eq.~(\ref{C21})
is zero.
In the calculations of multi-boson HBT correlation functions for the partially coherent
sources in the AMPT model, we substitute the factor $f(\mk_i-\mk_j,X)~(i,j=1,2,3,4,
i\neq j)$ in Eqs.~(\ref{C31})\,--\,(\ref{c41}) with $F(\mk_i-\mk_j,X)$, and assume
\begin{eqnarray}
&&F(\mk_i-\mk_j,X)=0, \hspace*{30mm}{\rm for}~\Delta z_{ij}<L_{ij\rm CZ}~{\rm and}~
\Delta r_{ij\rm T}<L_{ij\rm CT}, \cr
&&F(\mk_i-\mk_j,X)=f(\mk_i-\mk_j,X), \hspace*{25mm}
{\rm otherwise}, \hspace*{15mm}
\label{Fij}
\end{eqnarray}
where $\Delta z_{ij}=|z_i-z_j|$, $\Delta r_{ij \rm T}=\sqrt{(x_i-x_j)^2+(y_i-y_j)^2}$,~
$i,j=1,2,3,4,i\neq j$, and $L_{ij\rm CZ}$ and $L_{ij\rm CT}$ are the respective longitudinal and transverse coherent emission lengths.

In our previous work~\cite{WangYeZhang_PRC109_2024}, the coherent emission lengths are
proportional to the de Broglie wavelengths of the two pions, in which case the
proportionality coefficients should differ for different kinds of bosons, which are
produced at different evolution stages of the collision systems and have different multiplicities. For identical bosons emitted from a source with finite size, the emission
coherence is not only related to the bosons de Broglie wavelengths ($\sim h/k$) but also 
their density in the source. Considering that the emission coherence is related to the
boson density as well as the de Broglie wavelengths of the two bosons, we extend the
longitudinal and transverse coherent emission lengths in
Ref.~\cite{WangYeZhang_PRC109_2024} as
\begin{equation}
L_{ij \rm CZ}=a'_Z \langle D_Z\rangle \left[\frac{h}{k_{iZ}}+\frac{h}{k_{jZ}}\right],
\label{eq-LZ}
\end{equation}
and
\begin{equation}
L_{ij\rm CT}=a'_T \langle D_{T}\rangle \left[\frac{h}{k_{i\rm T}}+\frac{h}{k_{j
\rm T}}\right],
\label{eq-LT}
\end{equation}
where $i,j=1,2,3,4,i\neq j$; $a'_Z$ and $a'_T$ are longitudinal and transverse parameters,
respectively, which are universal constants for the collision system and can be determined
by comparing the model HBT results with experimental data; $k_{iZ}$ and $k_{iT}$ are the
respective boson longitudinal and transverse generation momenta.
Because the source properties in longitudinal and transverse directions are much different
in relativistic heavy-ion collisions, we introduced two independent coherent emission
lengths in longitudinal and transverse directions in Ref.~\cite{WangYeZhang_PRC109_2024},
which have different influences on the longitudinal and transverse HBT radii, to approach
the two-pion HBT results in the AMPT model with the experimental data
\cite{ALICE_PRC93_2016}.
In this paper, we assume that $\langle D_Z\rangle$ and $\langle D_T\rangle$ in
Eqs.~(\ref{eq-LZ}) and (\ref{eq-LT}) are the boson densities related to the longitudinal
and transverse coordinates, respectively,
\begin{equation}
\langle D_Z\rangle=\frac{M_b}{\langle r_Z\rangle},~~~~~~~~~~
\langle D_T\rangle=\frac{M_b}{\pi\langle r_T\rangle^2},
\end{equation}
where $M_b$ is the boson multiplicity, and $\langle r_Z\rangle$ and $\langle r_T\rangle$
are the respective averages of boson generation coordinates $r_Z=|z|$ and $r_T=\sqrt{x^2
+y^2}$ in the longitudinal and transverse directions.

Table I presents the identical pion and kaon densities $\langle D_Z\rangle$ and $\langle
D_T \rangle$ in Pb-Pb central collisions ($0\!<\!b\!<\!3.5$~fm) at $\sqrt{s_{NN}}=
2.76$~TeV in transverse momentum intervals $0\!<\!k_T\!<\!0.4$~GeV/c, $0.4\!<\!k_T
\!<\!1.0$~GeV/c, and $0\!<\!k_T\!<\!1.0$~GeV/c, calculated with $10^3$ events in the AMPT
model.

\begin{table*}[htb]
\begin{center}
\caption{Identical pion and kaon densities $\langle D_Z\rangle$ and $\langle D_T \rangle$
in Pb-Pb central collisions at $\sqrt{s_{NN}}=2.76$~TeV in different transverse momentum
intervals, calculated in the AMPT model.\vspace*{4mm}}
\begin{tabular}{c|ccc}
\hline\hline
~Pb-Pb@2.76\,TeV~&~~$0\!<\!k_T\!<\!0.4\,\text{GeV\!/\!c}$~~&
~~$0.4\!<\!k_T\!<\!1.0\,\text{GeV\!/\!c}$~~&
~~$0\!<\!k_T\!<\!1.0\,\text{GeV\!/\!c}$~~\\
\hline
$\pi$,~~$\langle D_Z\rangle$~[fm$^{-1}$]&213.744&286.782&248.840\\
$\pi$,~~$\langle D_T\rangle$~[fm$^{-2}$]&4.126&3.408&3.703\\
K,~~$\langle D_Z\rangle$~[fm$^{-1}$]&68.261&70.771&70.287\\
K,~~$\langle D_T\rangle$~[fm$^{-2}$]&1.770&0.912&1.017\\
\hline\hline
\end{tabular}
\label{Tab-D}
\end{center}
\end{table*}

It can be seen that the boson longitudinal densities in the higher transverse momentum
interval are greater than those in the lower transverse momentum interval and the boson
transverse densities in the higher transverse momentum interval are smaller than those in
the lower transverse momentum interval. It is because the average longitudinal coordinates
in the higher transverse momentum interval are smaller than those in the low transverse
momentum interval and the average transverse coordinates in the higher transverse
momentum interval are greater than those in the low transverse momentum interval.
In our calculations of multi-boson HBT correlations, we take the density values in the
interval $0\!<\!k_T\!<\!1.0$~GeV/$c$, which covers almost all generated pions and kaons
(99.9\% for pion and 99.8\% for kaon).

\section{Results of multiboson HBT correlation functions}
We investigate multi-boson HBT correlations as functions of the three- and four-boson
relative momentum variables
\cite{ALICE_PRC93_2016,Gangadharan_PRC92_2015,BaryRuZhang_JPG45_2018}
\begin{equation}
\label{q3}
Q_3=\sqrt{Q_{12}^2+Q_{23}^2+Q_{13}^2}
\end{equation}
and
\begin{equation}
\label{q4}
Q_4=\sqrt{Q_{12}^2+Q_{13}^2+Q_{14}^2+Q_{23}^2+Q_{24}^2+Q_{34}^2},
\end{equation}
where $Q_{ij}=\sqrt{-(p_i-p_j)^\mu(p_i-p_j)_\mu}$ ~($i,j=1,2,3,4$) is the invariant
relative momentum of the boson pair.
We investigate the multi-boson correlation functions in different intervals of the
transverse momenta,
\begin{equation}
K_{T3}=\frac{|\textbf{\emph{p}}_{1T}+\textbf{\emph{p}}_{2T}+\textbf{\emph{p}}_{3T}|}{3}
\end{equation}
and
\begin{equation}
K_{T4}=\frac{|\textbf{\emph{p}}_{1T}+\textbf{\emph{p}}_{2T}+\textbf{\emph{p}}_{3T}
+\textbf{\emph{p}}_{4T}|}{4}.
\end{equation}

\subsection{Multi-pion correlation functions}

\begin{figure}[htbp]
\hspace*{-5mm}
\includegraphics[scale=0.8]{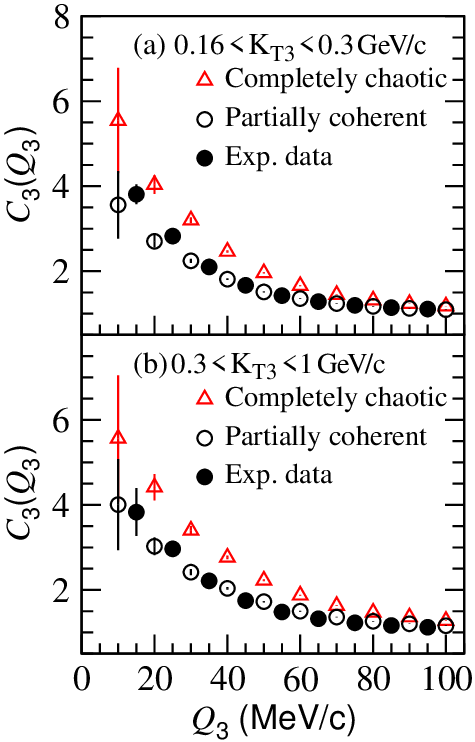}
\vspace*{-40mm}
\hspace*{5mm}
\includegraphics[scale=0.8]{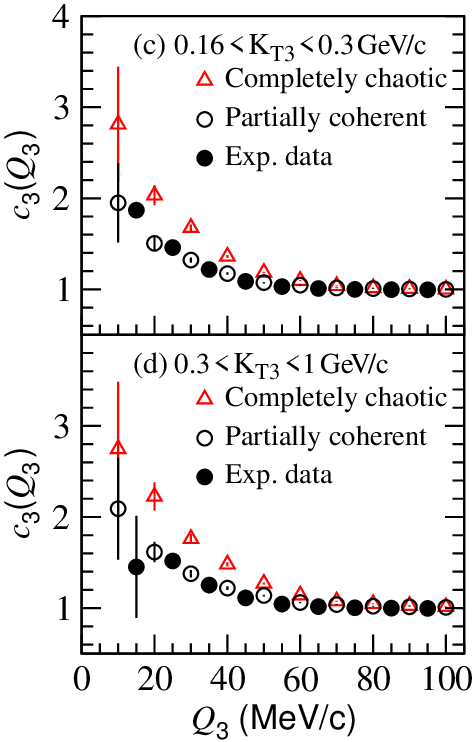}
\vspace*{40mm}
\caption{(Color online) Three-pion correlation functions $C_3(Q_3)$ and $c_3(Q_3)$ for
completely chaotic sources and partially coherent sources, respectively, in Pb-Pb central
collisions at $\sqrt{s_{NN}}=2.76$~TeV, calculated with the AMPT model. Experimental data
\cite{ALICE_PRC93_2016} are plotted for comparison. Results in transverse momentum
intervals $0.16\!<\!K_{T3}\!<\!0.3$GeV/$c$ and $0.3\!<\!K_{T3}\!<\!1$~GeV/$c$ are shown
in upper and lower panels, respectively.}
\label{zf_C3c3}
\end{figure}

Figure \ref{zf_C3c3} shows the three-pion correlation functions $C_3(Q_3)$ and $c_3
(Q_3)$ in Pb-Pb central collisions at $\sqrt{s_{NN}}=2.76$~TeV, in the low and high
transverse momentum intervals $0.16\!<\!K_{T3}\!<\!0.3$~GeV/$c$ and $0.3\!<\!K_{T3}\!<
\!1$~GeV/$c$, respectively. Here, the triangle and circle symbols refer to completely
chaotic and partially coherent sources, respectively, in the AMPT model with impact
parameter between 0 and 3.5~fm. Solid circles denote experimental data with centrality
0--5\% \cite{ALICE_PRC93_2016}. In the model calculations for the partially coherent
sources, we take the longitudinal and transverse parameters $a'_Z$ and $a'_T$ in
Eqs.~(\ref{eq-LZ}) and (\ref{eq-LT}) to be 0.0031~fm and 0.662~fm$^2$, respectively,
which are determined by comparing the model results with the experimental
data \cite{ALICE_PRC93_2016}. The values of densities $\langle D_Z\rangle$
and $\langle D_T\rangle$ are taken to be those in the momentum interval $0\!<\!k_T\!
<\!1.0$~Gev/$c$ in Table \ref{Tab-D}. The values of $a'_Z\langle D_Z\rangle$ and
$a'_T\langle D_T\rangle$ are consistent approximately with the corresponding parameters
$a_Z$ and $a_T$ in our previous work~\cite{WangYeZhang_PRC109_2024}, which investigated
the two-pion HBT radii as functions of pair momentum for the chaotic and partially coherent sources in the AMPT model in Pb-Pb collisions at $\sqrt{s_{NN}}=2.76$~TeV and compared them with the ALICE experimental data \cite{ALICE_PRC93_2016}.
To compare the multi-pion correlation functions with experimental data, we take the
pseudorapidity and transverse momentum cuts $|\eta|<0.8$ and $0.16<p_T<1.0$~GeV/$c$
\cite{ALICE_PRC93_2016} for pions in the AMPT model. The total number of model events
is $6\times 10^3$.

\vspace*{5mm}
\begin{figure}[htbp]
\hspace*{-5mm}
\includegraphics[scale=0.8]{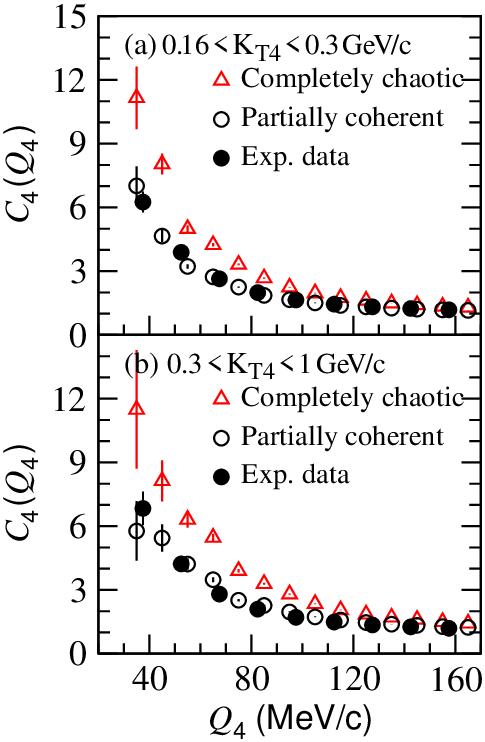}
\vspace*{-40mm}
\hspace*{5mm}
\includegraphics[scale=0.8]{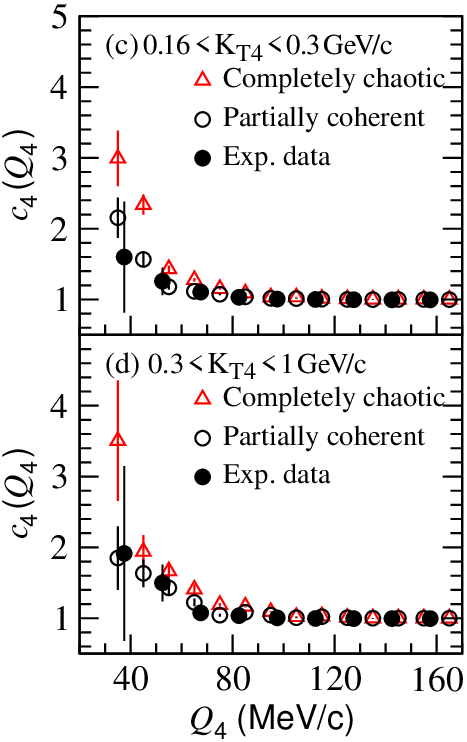}
\vspace*{40mm}
\caption{(Color online) Four-pion correlation functions $C_4(Q_4)$ and $c_4(Q_4)$ for
completely chaotic sources and partially coherent sources, respectively, in Pb-Pb
central collisions at $\sqrt{s_{NN}}=2.76$~TeV, calculated with the AMPT model. 
Experimental data \cite{ALICE_PRC93_2016} are plotted for comparison. 
Upper and Lower panels are
respective results in transverse momentum intervals $0.16\!<\!K_{T4}\!<\!0.3$~GeV/$c$
and $0.3\!<\!K_{T4}\!<\!1$~GeV/$c$.}
\label{zf_C4c4}
\end{figure}

\begin{figure}[htbp]
\hspace*{-5mm}
\includegraphics[scale=0.8]{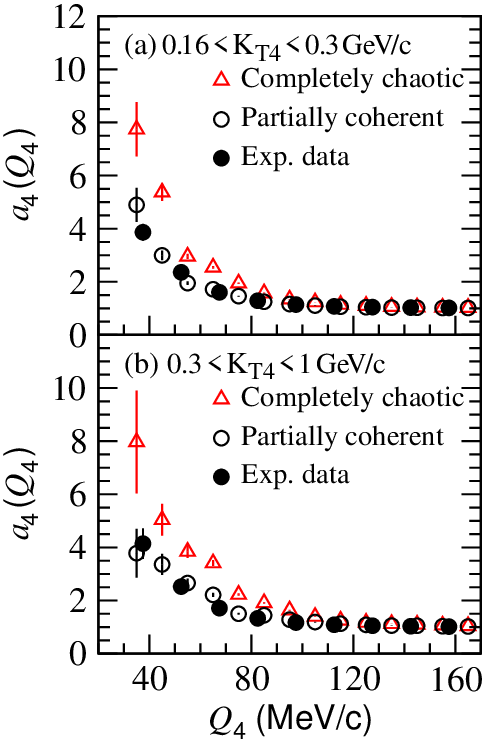}
\vspace*{-40mm}
\hspace*{5mm}
\includegraphics[scale=0.8]{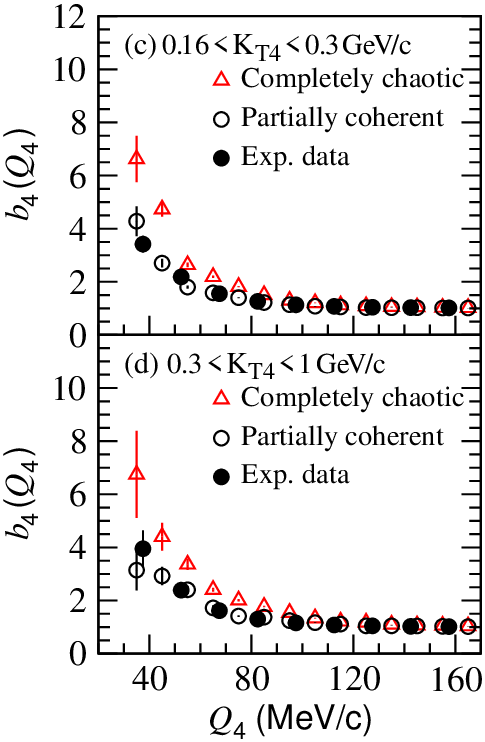}
\vspace*{40mm}
\caption{(Color online) Four-pion correlation functions $a_4(Q_4)$ and $b_4(Q_4)$ for
completely chaotic and partially coherent sources, respectively, in Pb-Pb central
collisions at $\sqrt{s_{NN}}=2.76$~TeV, calculated with the AMPT model.
Experimental data \cite{ALICE_PRC93_2016} are plotted for comparison.
Upper and lower panels are respective results in transverse momentum intervals
$0.16\!<\!K_{T4}\!<\!0.3$~GeV/$c$ and $0.3\!<\!K_{T4}\!<\!1$~GeV/$c$.}
\label{zf_a4b4}
\end{figure}

Figure \ref{zf_C4c4} shows the four-pion correlation functions $C_4(Q_4)$ and $c_4
(Q_4)$ in Pb-Pb central collisions at $\sqrt{s_{NN}}=2.76$~TeV, in the low and high
transverse momentum intervals $0.16\!<\!K_{T4}\!<\!0.3$~GeV/$c$ and $0.3\!<\!K_{T4}\!<
\!1$~GeV/$c$, respectively. As in Fig.~\ref{zf_C3c3}, the triangle and circle symbols
refer to completely chaotic and partially coherent sources, respectively, in the AMPT
model, and solid circles denote experimental data \cite{ALICE_PRC93_2016}.
Fig.~\ref{zf_a4b4} shows the four-pion correlation functions $a_4(Q_4)$ and $b_4(Q_4)$
for the completely chaotic and partially coherent sources in the AMPT model and the
experimental data in Pb-Pb central collisions at $\sqrt{s_{NN}}=
2.76$~TeV \cite{ALICE_PRC93_2016}, where symbol meanings are the same as in
Figs.~\ref{zf_C3c3} and \ref{zf_C4c4}.

\vspace*{5mm}

From Figs.~\ref{zf_C3c3}, \ref{zf_C4c4}, and \ref{zf_a4b4}, one can see that the three-
and four-pion correlation functions with various cumulants for partially coherent
sources in the AMPT model are all consistent with the experimental data in Pb-Pb central
collisions at $\sqrt{s_{NN}}=2.76$~TeV \cite{ALICE_PRC93_2016}. However, the results of
the three- and four-pion correlation functions for completely chaotic sources are higher
than the experimental data \cite{ALICE_PRC93_2016}.
This consistency that all of the multi-pion correlation functions with experimental data
may indicate that the partially coherent sources constructed in the AMPT model describe
well the emission coherence of identical pions in the heavy-ion collisions.

\subsection{Multi-kaon correlation functions}

In Eqs.~(\ref{eq-LZ}) and (\ref{eq-LT}), $a'_Z$ and $a'_T$ are considered as two universal
constant parameters related only to the collision systems. They are determined in subsection
III A to be 0.0031~fm and 0.662~fm$^2$, respectively, by comparing the model results with
experimental data of three- and four-pion correlation functions. We assume here that these
parameter values are also suitable for the kaon sources in the collisions, and investigate
the three- and four-kaon correlation functions in Pb-Pb central collisions
at $\sqrt{s_{NN}}=2.76$~TeV in the AMPT model.

\begin{figure}[htbp]
\hspace*{-5mm}
\includegraphics[scale=0.8]{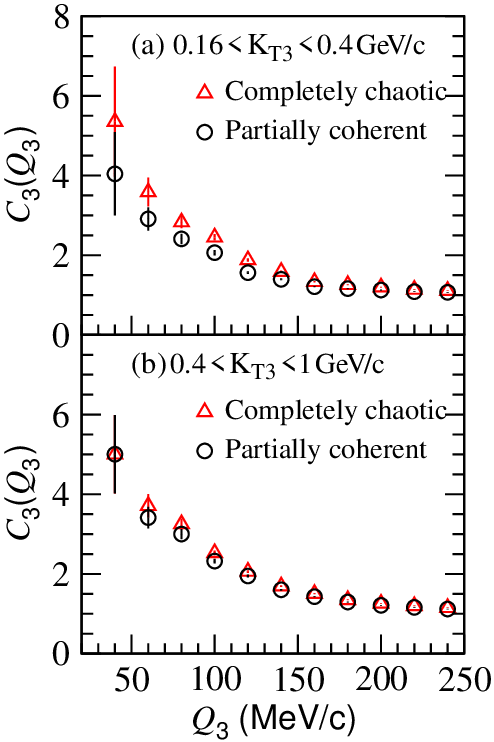}
\vspace*{-40mm}
\hspace*{5mm}
\includegraphics[scale=0.8]{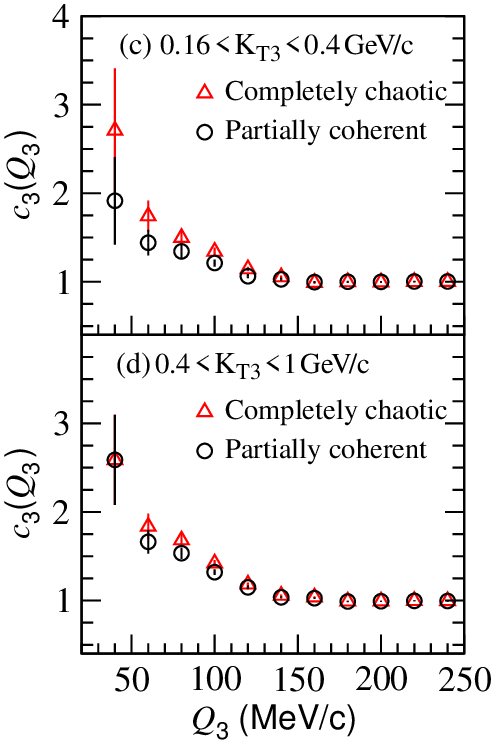}
\vspace*{40mm}
\caption{(Color online) Three-kaon correlation functions $C_3(Q_3)$ and $c_3(Q_3)$ for
completely chaotic and partially coherent sources, respectively, in Pb-Pb central collisions
at $\sqrt{s_{NN}}=2.76$~TeV, calculated with the AMPT model. Upper and lower panels, respectively,
are results in transverse momentum intervals $0.16\!<\!K_{T3}\!<\!0.4$ GeV/$c$ and $0.4\!<\!
K_{T3}\!<\!1$~GeV/$c$.}
\label{zf_C3c3K}
\end{figure}

Figure \ref{zf_C3c3K} shows the three-kaon correlation functions $C_3(Q_3)$ and $c_3
(Q_3)$ in Pb-Pb central collisions at $\sqrt{s_{NN}}=2.76$~TeV, in low and high
transverse momentum intervals $0.16\!<\!K_{T3}\!<\!0.4$~GeV/$c$ and $0.4\!<\!K_{T3}\!<
\!1$~GeV/$c$, respectively. Triangle and circle symbols, respectively, are for completely
chaotic and partially coherent sources in the AMPT model. We take the kaon sample with
the same pseudorapidity and transverse momentum cuts $|\eta|<0.8$ and $0.16<p_T<
1.0$~GeV/$c$ as in pion sample \cite{ALICE_PRC93_2016} for comparison. The total event
number for the multi-kaon correlation calculations is $4\times 10^4$. One can see that
the kaon correlation functions are wider than those of the pion (see Fig.~\ref{zf_C3c3}),
because the kaon source sizes are smaller. The differences between the kaon correlation
functions of the completely chaotic and partially coherent sources are smaller than those
of the pion sources. This means the kaon sources having lower coherence than the pion
sources.

\vspace*{5mm}
\begin{figure}[htbp]
\hspace*{-5mm}
\includegraphics[scale=0.8]{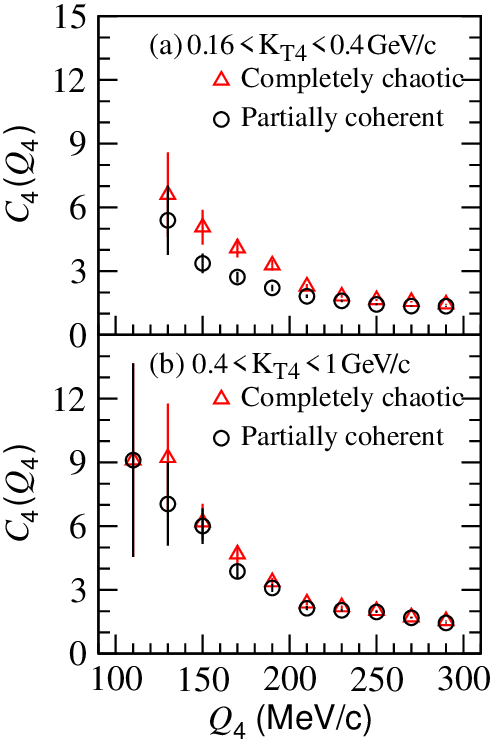}
\vspace*{-40mm}
\hspace*{5mm}
\includegraphics[scale=0.8]{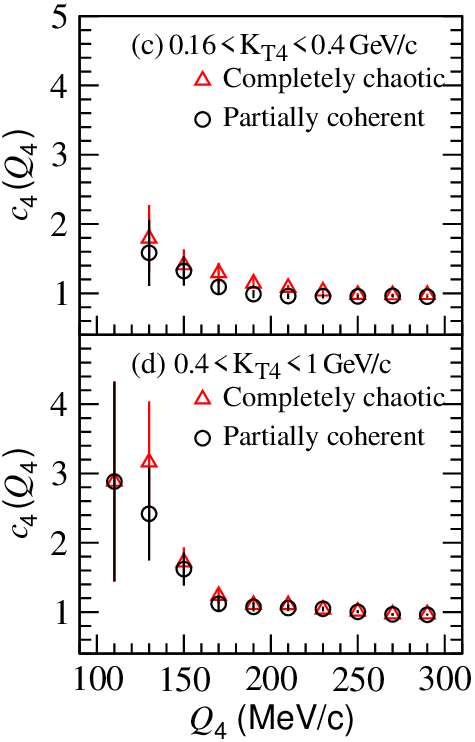}
\vspace*{40mm}
\caption{(Color online) Four-kaon correlation functions $C_4(Q_4)$ and $c_4(Q_4)$
for completely chaotic and partially coherent sources, respectively, in Pb-Pb central
collisions at $\sqrt{s_{NN}}=2.76$~TeV, calculated with the AMPT model. Upper and lower
panels are results in transverse momentum intervals $0.16\!<\!K_{T4}\!<\!0.4$ GeV/$c$
and $0.4\!<\!K_{T4}\!<\!1$~GeV/$c$, respectively.}
\label{zf_C4c4K}
\end{figure}

\vspace*{5mm}
\begin{figure}[htbp]
\hspace*{-5mm}
\includegraphics[scale=0.8]{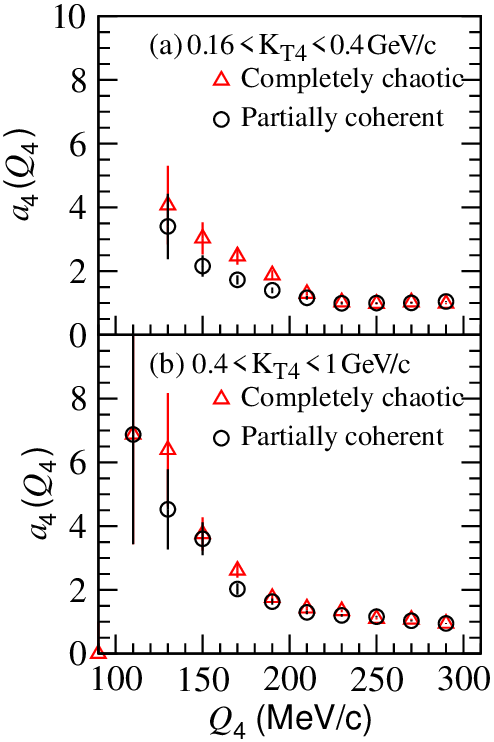}
\vspace*{-40mm}
\hspace*{5mm}
\includegraphics[scale=0.8]{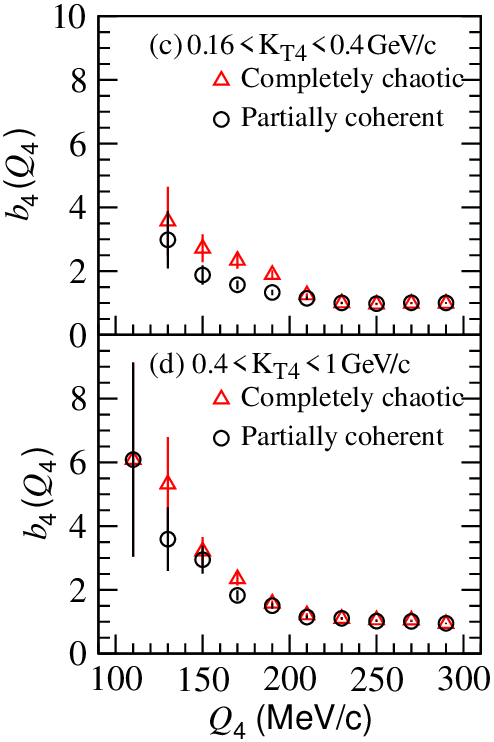}
\vspace*{40mm}
\caption{(Color online) Four-kaon correlation functions $a_4(Q_4)$ and $b_4(Q_4)$
for completely chaotic and partially coherent sources, respectively, in Pb-Pb central
collisions at $\sqrt{s_{NN}}=2.76$~TeV, calculated with the AMPT model. Upper and lower
panels are results in transverse momentum intervals $0.16\!<\!K_{T4}\!<\!0.4$ GeV/$c$
and $0.4\!<\!K_{T4}\!<\!1$~GeV/$c$, respectively.}
\label{zf_a4b4K}
\end{figure}

Figure \ref{zf_C4c4K} shows the four-kaon correlation functions $C_4(Q_4)$ and $c_4(Q_4)$,
and Figure~\ref{zf_a4b4K} shows the four-kaon correlation functions $a_4(Q_4)$ and
$b_4(Q_4)$ in Pb-Pb central collisions, at $\sqrt{s_{NN}}=2.76$~TeV. It can be seen that
the differences of $C_4(Q_4)$, $a_4(Q_4)$, and $b_4(Q_4)$ between the completely chaotic
and partially coherent sources in the lower $K_{T4}$ interval are greater than those in
the higher $K_{T4}$ interval. The data at the smallest $Q_4$ in the higher $K_{T4}$
interval are from the contributions of the sampled four kaons with small angles among
their transverse momenta.

\subsection{Multiboson correlation function intercepts and equivalent coherence
parameters of partially coherent sources}

In this study, we introduce the longitudinal and transverse coherent-emission lengths to
construct the partially coherent sources of identical bosons in the AMPT model. These
two coherent-emission lengths are particle momentum and density dependent. To investigate
the influence of coherent emission on multi-boson HBT correlations, we introduce the
intercept parameters of the three- and four-boson correlation functions $\lambda_3$ and
$\lambda_4$, respectively, and write the pure three- and four-boson correlation functions
for the partially coherent sources as
\begin{equation}
c_3(\mk_1,\mk_2,\mk_3)=1+\lambda_3\, R_3(\mk_1,\mk_2,\mk_3),
\label{C3R3}
\end{equation}
\begin{equation}
c_4(\mk_1,\mk_2,\mk_3,\mk_4)=1+\lambda_4\, R_4(\mk_1,\mk_2,\mk_3,\mk_4),
\label{C4R4}
\end{equation}
where $R_3(\mk_1,\mk_2,\mk_3)$ and $R_4(\mk_1,\mk_2,\mk_3,\mk_4)$ are pure three- and
four-boson correlators, respectively, which are functions of the respective relative
momenta $Q_3$ and $Q_4$. At $Q_3=0$ and $Q_4=0$, we have $\lambda_3=2$ and $\lambda_4=6$ for
completely chaotic sources and $\lambda_3=\lambda_4=0$ for completely coherent sources
[ see Eqs. (\ref{c31}) and (\ref{c41}) ]. Because of the absence of sufficient statistics
of the correlators near zero relative momenta $Q_3$ and $Q_4$, it is hard to extract the
intercepts directly from the correlation function data.

Based on our calculations for the correlation functions of the partially coherent sources
in the AMPT model, the intercepts can be written as
\begin{equation}
\lambda_3=2\frac{n_{3\chi}}{n_{3t}}, ~~~~~~~~~~~~\lambda_4=6\frac{n_{4\chi}}{n_{4t}},
\end{equation}
where $n_{3\chi}$ is the number of the boson triplets in which all pairs satisfied
$F(\mk_i-\mk_j,X)=f(\mk_i-\mk_j,X)$ $(i,j=1,2,3,i\ne j)$ in Eq.~(\ref{Fij}) in the
three-boson samples, $n_{4\chi}$ is the number of the boson quadruplets in which all
pairs satisfied  $F(\mk_i-\mk_j,X)=f(\mk_i-\mk_j,X)$ $(i,j=1,2,3,4,i\ne j)$
in Eq.~(\ref{Fij}) in the four-boson samples, and $n_{3t}$ and $n_{4t}$ are the total
numbers of the three- and four-boson samples, respectively.

For a static partially coherent boson-emitting source with the same Gaussian spatial
distribution for chaotic and coherent emissions and a constant ratio $\gamma$ of the
coherent emission contribution $b_c$ to the chaotic emission contribution $b_{\chi}$
[~$\gamma=(b_c/b_{\chi})$~], one has with the diagram technique \cite{Liu_PRC_86}
\begin{equation}
c_3(\mk_1,\mk_2,\mk_3)=1+2\xi_{_G}\, e^{-(q_{12}^2+q_{13}^2+q_{23}^2)R^2\!/2},
\label{C3G3}
\end{equation}
\begin{eqnarray}
&&c_4(\mk_1,\mk_2,\mk_3,\mk_4)=1+2\eta_{_G}\, \big[e^{-(q_{12}^2+q_{23}^2+q_{34}^2+q_{41}^2)
R^2\!/2}\nonumber\\
&&\hspace*{8mm}+\,e^{-(q_{12}^2+q_{24}^2+q_{43}^2+q_{31}^2)R^2\!/2}
+e^{-(q_{13}^2+q_{32}^2+q_{24}^2+q_{41}^2)R^2\!/2}\big],
\label{C4G4}
\end{eqnarray}
where, $q_{ij}=|\mk_i-\mk_j|~(i,j=1,2,3,4)$, $R$ is the standard variance of the Gaussian
distribution, and
\begin{equation}
\xi_{_G}=\frac{1+3\gamma}{(1+\gamma)^3},~~~~~~~~\eta_{_G}=\frac{1+4\gamma}{(1+\gamma)^4}.
\end{equation}

Letting $\lambda_3=2\xi_{_G}$ and $\lambda_4=6\eta_{_G}$, we can obtain the equivalent
coherent fractions, $C\!F_m=b_c/(b_c+b_{\chi})$, $m=3,4$, with intercepts $\lambda_3$ and
$\lambda_4$ for the partially coherent sources in the AMPT model. The values of equivalent
coherent fractions $C\!F_m$ ($m=2,3,4$) provide another description of the source coherence.

Table \ref{lmCF} presents the intercept values of pion and kaon multi-particle correlation
functions in the low and high transverse momentum intervals for partially coherent
sources in Pb-Pb central collisions at $\sqrt{s_{NN}}=2.76$~TeV/$c$ in the AMPT model.
The statistical errors of the intercepts are proportional to $(n_{\chi}^{-1/2}+n_t^{-1/2})$,
which are insignificant for large $n_{\chi}$ and $n_t$. The corresponding $C\!F$ results
are also presented in Table \ref{lmCF}.

\begin{table}[htb]
\begin{center}
\caption{Intercepts of multi-boson correlation functions ($\lambda_3$, $\lambda_4$) and
equivalent coherent fractions ($C\!F_3$, $C\!F_4$) for partially coherent sources in
Pb-Pb central collisions at $\sqrt{s_{NN}}=2.76$~TeV in the AMPT model. \vspace*{4mm}}
\begin{tabular}{c|cc|cc}
\hline\hline
Pb-Pb@2.76\,TeV&~~$\lambda_3$~&~~$C\!F_3$~~&~~$\lambda_4$~&~~$C\!F_4$~~\\
\hline
~$\pi$, $0.16\!<\!K_{Tm}\!<\!\!0.3\,\text{GeV\!/\!c}$&~~0.851~~&~~0.550~~&~~2.049~~&~~0.481~\\
~$\pi$,\, $0.3\!<\!K_{Tm}\!<\!1.0\,\text{GeV\!/\!c}$&~~0.934~~&~~0.522~~&~~2.335~~&~~0.451~\\
~K, $0.16\!<\!K_{Tm}\!<\!\!0.4\,\text{GeV\!/\!c}$&~~1.409~~&~~0.360~~&~~5.720~~&~~0.094~\\
~K, $0.4\!<\!K_{Tm}\!<\!1.0\,\text{GeV\!/\!c}$&~~1.740~~&~~0.226~~&~~5.751~~&~~0.088~\\
\hline\hline
\end{tabular}
\label{lmCF}
\end{center}
\end{table}

It can be seen from Table \ref{lmCF} that the intercepts of the multi-pion and multi-kaon
correlation functions of the partially coherent sources are higher in the high transverse
momentum intervals than those in the low transverse momentum intervals, and similarly
the equivalent coherent fractions are smaller. This is because the particle transverse
de Broglie wavelengths are small at high transverse momenta. The intercept results of
multi-kaon correlation functions are larger than those of multi-pion correlation functions
because the kaon densities are less than the pion densities ( see table \ref{Tab-D} ).
The inconformity of the equivalent coherent fraction $C\!F_3$ with $C\!F_4$ occurs because
the partially coherent sources in the AMPT model are different from those of the static
Gaussian model.

\section{Summary and conclusion}
In this study. we investigated multi-pion and multi-kaon HBT correlations for partially
coherent and completely chaotic sources in Pb-Pb central collisions at $\sqrt{s_{NN}}=2.76
$~TeV in the AMPT model. The longitudinal and transverse coherent emission lengths of the
partially coherent sources are proportional to the boson de Broglie wavelengths as well
as to the boson densities, and the proportionality coefficients are assumed to be universal
constants both for pion and kaon partially coherent particle-emitting sources.
We calculated the intercepts of the multi-pion and multi-kaon correlation functions of
the partially coherent sources in the AMPT model, and investigated the influence of boson
coherent emission on multi-boson HBT correlations.

It was found that the multi-boson correlation functions of partially coherent sources
were lower than those of completely chaotic sources. All of the three- and four-pion
correlation functions of the partially coherent sources are consistent with experimental
data. This consistency may indicate that the partially coherent source constructed in the
AMPT model well describes the emission coherence of identical pions in heavy-ion collisions.
Multi-kaon correlation functions are wider than multi-pion correlation functions because the
kaon sources are smaller than pion sources.
The intercepts of multi-boson correlation functions exhibit the influence of the source
coherence on HBT correlations for partially coherent sources. The intercepts of the three-
and four-kaon correlation functions of partially coherent sources are greater than those of
three- and four-pion correlation functions, because low kaon densities lead to the smaller
kaon coherent-emission lengths than the pion coherent-emission lengths. The intercepts of
multi-pion and multi-kaon correlation functions of partially coherent sources in high
transverse momentum intervals are higher than those in low transverse momentum intervals,
because particle de Broglie wavelengths are small at high momenta. Our study also indicates
that the partially coherent sources constructed in the AMPT model are different from those
of a static Gaussian model.
In this work, the coherent emissions in longitudinal and transverse directions are independent and have sharp boundaries. In future work, investigating their relation
and considering more realistic coherent emission regions, for instance, with Gaussian distributions, are of interest.
In addition, an investigation of two- and multi-boson HBT correlations for more collision
systems in microscopic cascade models will be of interest.

\begin{acknowledgements}
We thank Zi-Wei Lin for useful discussions and suggestions.
This research was supported by the National Natural Science Foundation of
China under Grant No. 12175031. We thank LetPub (www.letpub.com.cn) for its
linguistic assistance during the preparation of this manuscript.
\end{acknowledgements}

\end{document}